\documentstyle{llncs}
\input epsf

\begin{document}
\title{Image Synthesis and Style Transfer}

\author{Somnuk Phon-Amnuaisuk$^{1,2}$}
\institute{Media Informatics Special Interest Group, \\
$^{1}$Centre for Innovative Engineering, Universiti Teknologi Brunei,\\
$^{2}$School of Computing \& Informatics, Universiti Teknologi Brunei.
\email{somnuk.phonamnuaisuk@utb.edu.bn}
}

\maketitle

\begin{abstract}
Affine transformation, layer blending, and artistic filters are popular processes that graphic designers employ to transform pixels of an image to create a desired effect. Here, we examine various approaches that synthesize new images: pixel-based compositing models and in particular, distributed representations of deep neural network models.  This paper focuses on synthesizing new images from a learned representation model obtained from the VGG network. This approach offers an interesting creative process from its distributed representation of information in hidden layers of a deep VGG network i.e., information such as contour, shape, etc. are effectively captured in hidden layers of neural networks. Conceptually, if $\Phi$ is the function that transforms input pixels into distributed representations of VGG layers ${\bf h}$, a new synthesized image $X$ can be generated from its inverse function, $X = \Phi^{-1}({\bf h})$. We describe the concept behind the approach, present some representative synthesized images and style-transferred image examples.
\end{abstract}

\subsubsection*{Keywords}
Image synthesis, Creative generative process, Distributed representations, VGG network

\section{Introduction}
Computer has been extensively employed to process 2D vector graphics and raster graphics. Human designers apply various techniques such as cropping, compositing, transforming e.g., scaling, rotating, and applying various visual effects using filtering techniques. We may further describe the graphic content processing of raster graphics into two main approaches: (i) manually or algorithmically composing existing images into a new graphical content; and (ii) creating a new graphical content using generative models that are algorithmically crafted\footnote{One of the pioneers in this area is Harold Cohen \texttt{aaronshome.com/aaron/index.html}} \cite{span12,nina15,mandelbrot83,lindenmayer90}; or learning from image examples, e.g., deep dream\footnote{\texttt{https://github.com/google/deepdream}}. The first approach is the more popular contemporary techniques where graphic designers employ off-the-shelf graphics authoring tools to create new content from existing images. In this work, we are particularly interested in the second approach where a new graphical content is synthesized at pixel level using a generative model learnt from examples\footnote{In this paper, the terms \emph{synthesis} and \emph{generate} are often used interchangably.}. This approach has received much interest in recent years.

Recent advances in deep neural networks have shed some interesting ideas on the knowledge representation and learning issues \cite{schmidhuber15,lecun15}.  It is found that hidden nodes of a deep convolution neural networks (CNNs) trained with audio or visual stimuli could represent the basic fundamental frequency of sound or basic visual patterns \cite{zeiler13,zhou17}. 
These ideas motivate us to look into the synthesis of an image from a distributed representation. In this paper, we explore a generative model that synthesizes new images using a distributed representation obtained from VGG16 network \cite{simonyan15}. More information about the VGG16 deep convolution neural network will be discussed in Section \ref{syndist}.

The rest of the paper is organized into the following sections: Section 2 discusses the background and some representative related works; Section 3 discusses our approach and gives the details of the techniques behind it; Section 4 provides a critical discussion of the output from the proposed approach; and finally, the conclusion and further research are presented in Section 5.

\section{Related Works}
A two dimensional image is a projection of the three dimensional world onto a 2D plane, with a primitive unit as pixel. This process abstracts the world to only pixel-intensity and colour. 
Features such as histograms, edges, contours, corner points, object skeleton after erosion, lines regions, Fourier coefficients, convolution filter coefficients, etc. have been extensively exploited to infer the original information of the world from pixel information. 
Much research in computer vision has been poured into object recognition and understanding of scenic content from features derived from pixel information. It is interesting to explore whether it would be possible to reverse the process and synthesize an image using the features mentioned.

Abstract patterns have been commonly generated using mathematical functions \cite{kalajd08}.
Random patterns, fractal and abstract patterns can normally be expressed using rather short finite length programs. Although many insights have been observed and formulated in the graphic design domain, such as the rule of third and the golden ratio,
there is still a big gap in our understanding of how a non-abstract image can be automatically generated using a computer program i.e., creating an outdoor scenery on a blank canvas.
Due to its complexity, researchers have often choosen to explicitly describe the generative process as a computer program \cite{span12} or describing a sequence of image processing operations to be applied to existing images, e.g., \emph{non-photo realistic rendering} (NPR) \cite{strothotte02}. 

In the more popular contemporary approach, a new image is created by modifying original image using transformation, artistic filters and various composition tactics. The existing images will go through various processes, where their color, shapes, texture are modified and then re-composited into a new image \cite{spanazhan14}. In this style, the content is consciously modified by a human designer who asserts all extraneous information to the composited content. Various commercial graphics software packages have been designed to assist human designers in this approach. Humans employ both top-down and bottom-up creative processes in this approach. 

One of the early works attempting to automate the above creative process is from \cite{hertzmann01}, the authors propose the concept named \emph{image analogies}. Given images $A, A'$ and $B$ where $f(A) \rightarrow A'$, $f$ is a function that transform $A$ to $A'$ e.g., $f$ could be an artistic filter function. The authors show that the transforming function $f$ could be learnt from an example pair $(A,A')$ and then apply the learnt function to generate a new output $B'$. This approach provides an interesting \emph{style transferred} process.
 
With the recent advances in \emph{deep learning}, it has been shown that the convolution filters of a convolution neural network (CNN) exhibit important characteristics of the way our brain responds to visual patterns; e.g., simple patterns such as lines, dots, colors that emerge in early layers and complex patterns such as textures, and compound structures that emerge in deeper layers \cite{zeiler13,zhou17}. 

The weights of CNN can be seen as a function that re-represents an input image $I$ in hidden nodes ${\bf h}$ of CNN $f: I \mapsto {\bf h}$. Given an input image to a trained CNN, information residing in the trained CNN can be transferred to the input image by enhancing the activation signal on the hidden nodes that responds strongly to the input image. The gradient of the hidden nodes with respect to the input image could be used to modify the input image. Iteratively repeating the process will enhance the features strongly correlated with those hidden nodes e.g., abstract geometrical patterns and complex patterns that have been learnt by the network in the resulting image. Google deep dream is one of the influential works that employs this technique to generate images from its network trained with images of cats and dogs. Feeding a new input image such as the sky, the network generates a kind of hallucinative flavour by embedding parts of cats and dogs to the output images. This sparks many subsequent works in image synthesis and style transfer \cite{gatys16}.

\section{Formalizing the Pixel-based Image Synthesis Process}

\subsection{Composite Models}
A composite model generates a new image by modifying existing pixels information through the modification of pixel intensity, through affine transformation, convolution, or compositing pixels from various sources. We highlight some important operations in the categories below.\subsubsection{Convolutional Filters:}
Let $X_i^{r \times c}$ be a matrix of pixels of image $i$ where $x_i(r,c)$ is the pixel intensity. 
We can express convolution operation between image $i$ and a convolution kernel $K$ as
\begin{equation} \label{convol}
X' \leftarrow   K*X_i
\end{equation}
where $*$ is a convolution operator and $X'$ is the output image. 
\subsubsection{Layers Blending:}
We can express the composition between image $i$ and $j$ as
\begin{equation} \label{blend}
X' \leftarrow  \alpha X_i + (1-\alpha) X_j 
\end{equation}
where $\alpha \in [0,1]$ controls the blending percentage of images $i$ and $j$.  Various operations such as affine transformation, intensity/color manipulation, and convolution operations can be expressed as a sequence of these basic processes.

\begin{figure}[!ht]
\begin{center}\leavevmode
\epsfxsize=12cm
\epsfbox{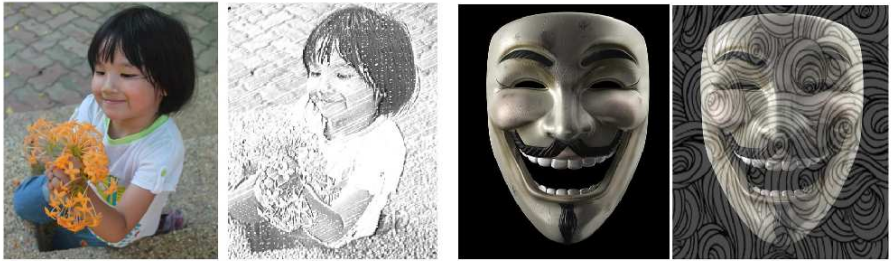}
\end{center}
\caption{Composite models: the image in column two is generated by applying a filter to emulate the pencil sketch effect (see \cite{spanazhan14}). The image in column three is generated by compositing two images together.}
\label{compsynim1}
\end{figure}
\begin{figure}[!ht]
\begin{center}\leavevmode
\epsfxsize=12cm
\epsfbox{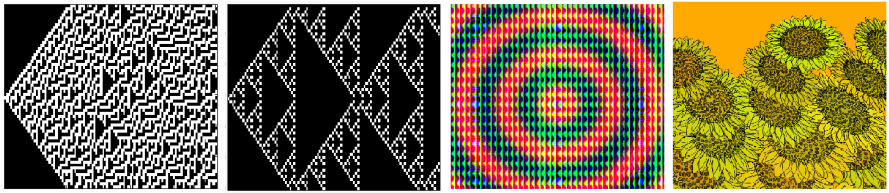}
\end{center}
\caption{Automated compositing models: the images in columns one and two are generated using cellular automata \cite{wolfram84} with rule 30 and rule 90 respectively. The image in column three is generated using a compact mathematical formula while the image in column four is generated using an elaborate procedure that draws sunflower seeds, petals and their compositions (see \cite{span12}).}
\label{comsynim2}
\end{figure}
Figure \ref{compsynim1} shows the output from composite approach. Columns one and three show the original image and columns two and four shows the processed images. The output in columns two and four are generated using Eqs. \ref{convol} and \ref{blend} respectively. The generated images closely resemble the original images. In this style, the variation is manually controlled by a human graphic designer. 
The process above could be automated using computers. Weaving many small generative units together into algorithms, complex processes such as drawing sunflower seeds and petals could be automated \cite{span12}. Figure \ref{comsynim2} shows a more complex composite model generated by computer programs.

\subsection{Learned Distributed Representation Models}
Let $X^{r \times c}$ be a matrix of input pixels and $G_h^{r \times c}$ be a gradient matrix computed from hidden nodes parameters. Conceptually speaking, the generative model that optimizes the activations of hidden nodes can be expressed as:
\begin{equation}
X \leftarrow  X + \delta G_h
\end{equation}
where $\delta$ is the step size, that is, an image $X$ gradually transforms according to the added gradient. Images produced in this style are interesting but the contents are random in nature since the generated image is dependent on the input pixels and the contents learnt by the network. There is no means to control the relationships among the components in the generated contents.

A generative model can be viewed as a function. If this function can be precisely determined, then, given an input image, a synthesized image can be computed and vice versa, given a synthesized image, an original image can be computed using its reverse function. Here the functionality of this generative function is emulated in an artificial neural network architecture using the process explained below. 

Let $\Phi$ be the function that transforms input pixels into a feature vector {\bf h} derived from hidden nodes (e.g., weights or activations) distributed in different layers of the network ${\bf h} \leftarrow \Phi(X)$. Conceptually, the original image can be reproduced from a distributed representation of ${\bf h}$ \cite{mahendran15}: 
\begin{equation}
X \leftarrow  \Phi^{-1}({\bf h})
\end{equation}  
This concept can be extended to the generation of $X$ from weighted ${\bf h}_i$ of multiple transform functions, $i$.
\begin{equation}
X \leftarrow  \sum_i w_i\Phi_i^{-1}({\bf h}_i)
\label{eq2}
\end{equation}

\section{Synthesizing Images from Distributed Representations}
\label{syndist}
\begin{figure}[!ht]
\begin{center}\leavevmode
\epsfxsize=13cm
\epsfbox{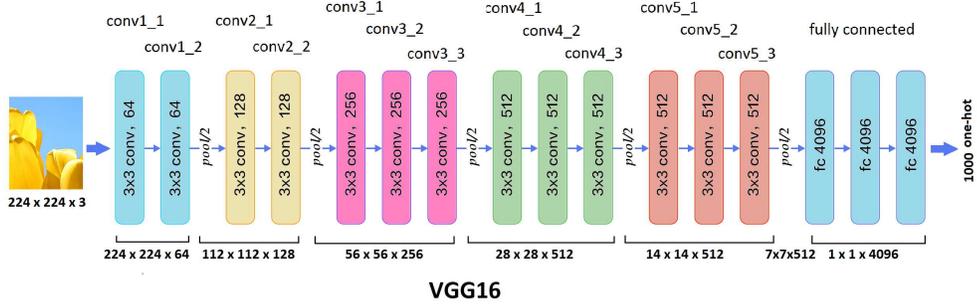}
\end{center}
\caption{The graphical abstract representation of the VGG16.}
\label{vgg16}
\end{figure}

We exploit the VGG deep neural network in our image synthesis task. VGG is the acronym for \emph{Visual Geometry Group from the University of Oxford}. The group has released two fully trained deep convolution neural networks, namely VGG16 and VGG19, to the public\footnote{see www.robots.ox.ac.uk/~vgg/research/very\_deep}. Here, we experiment with image synthesis using parameters read from convolution layers of VGG16 network.  Fig. \ref{vgg16} shows the architecture of VGG16 network. Each block represents a hidden layer, for example, \emph{3$\times$3 conv, 64} denotes 64 convolution filters of size 3$\times$3.


Feeding an image $X^{r \times c}_i$ to the VGG network, we observe the activations in the all hidden convolution layers $l$, ${\bf h}^l_i$. In other words, activations in all hidden layers represent the pixel information of the input image and, conceptually, a copy of $X$ should be reproducible by reverse the process $X'_i \leftarrow  \Phi^{-1}({\bf h}^l_i)$. In this work, instead of analytically solving of a function $\Phi$ and its inverse  $\Phi^{-1}$, we adopt an optimization method that gradually adjusts the activation of hidden nodes to the desired ${\bf h}^l_i$.

In \cite{gatys15}, the authors approach the style transfer task by mnimizing two kinds of loss functions: content loss and style loss. 
Let $P^l \in {\mathcal R}^{N_l \times M_l}$, $S^l \in {\mathcal R}^{N_l \times M_l}$ and $F^l \in {\mathcal R}^{N_l \times M_l}$ be three matrices derived from the layer $l$ of the VGG network fed with content image, style image and noise input respectively. $N_l$ denotes the number of feature maps and $M_l$ denotes the size of the feature map. The style loss and content loss are defined as follows:
\begin{equation} \label{style}
{\mathcal L}_{style} = \frac{1}{4{N_l}^2 {M_l}^2} \sum_{i=1}^{N_l} \sum_{j=1}^{N_l} (G_{ij}^l - A_{ij}^l)^2
\end{equation}
\begin{equation} \label{content}
{\mathcal L}_{content} = \frac{1}{2}\sum_{i=1}^{N_l} \sum_{j=1}^{M_l} (F_{ij}^l - P_{ij}^l)^2
\end{equation}
\begin{equation} \label{gram1}
A_{ij}^l= \sum_{k=1}^{M_l} S_{ik}^l S_{jk}^l 
\end{equation}
\begin{equation} \label{gram2}
G_{ij}^l= \sum_{k=1}^{M_l} F_{ik}^l F_{jk}^l
\end{equation}
where $A_{ij}^l$ and $G_{ij}^l$ is the Gram matrix which is calculated from the inner product of $S_{ik}^{l} S_{jk}^l$ and $F_{ik}^{l} F_{jk}^l$ respectively.

\subsection{Synthesized Images}
\begin{figure}[!ht]
\begin{center}\leavevmode
\epsfxsize=12cm
\epsfbox{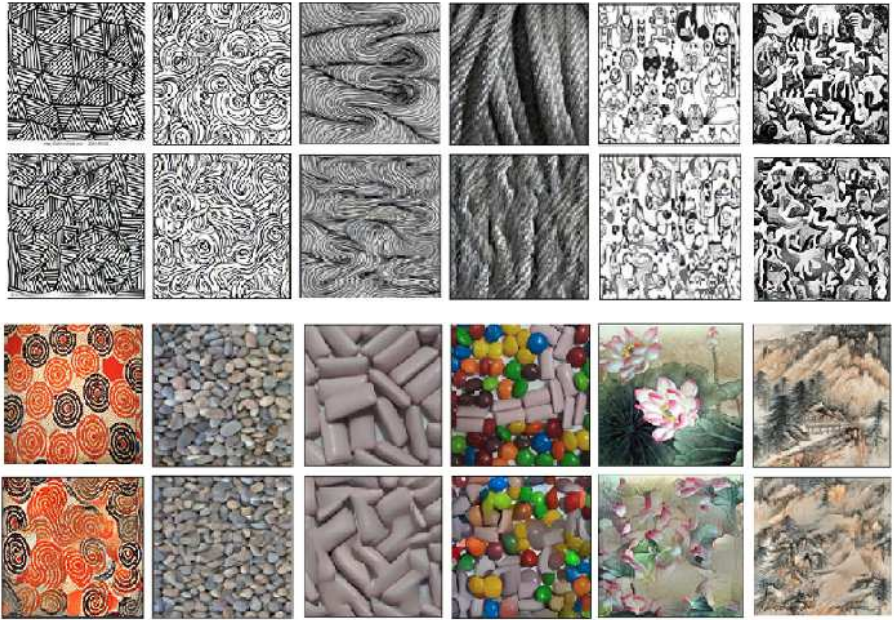}
\end{center}
\caption{Synthesized images using equation \ref{style} display good similarities at the local level but relationships among image components over a spatial distance are lost (see synthesized images in the last two columns).}
\label{resultsynim1}
\end{figure}

Figure \ref{resultsynim1} presents twelve images in two groups. The first and the third rows are the original images; the second and the third rows are the synthesized images based on the original images. These twelve images are chosen to highlight the characteristics of synthesized images using Eq \ref{style}. Four patterns (the two-rightmost columns) are chosen to display strong long range dependency. It is clear that all synthesized images successfully capture local dependency using the ${\mathcal L}_{style}$ loss. However, long range dependency is not successfully captured. It is clear from the synthesized images in the two rightmost columns that although the synthesized patterns seem to capture the style (local dependency) in general, the dependency among components in a longer distance are lost and a lot of information is missing.

\begin{figure}[!ht]
\begin{center}\leavevmode
\epsfxsize=12cm
\epsfbox{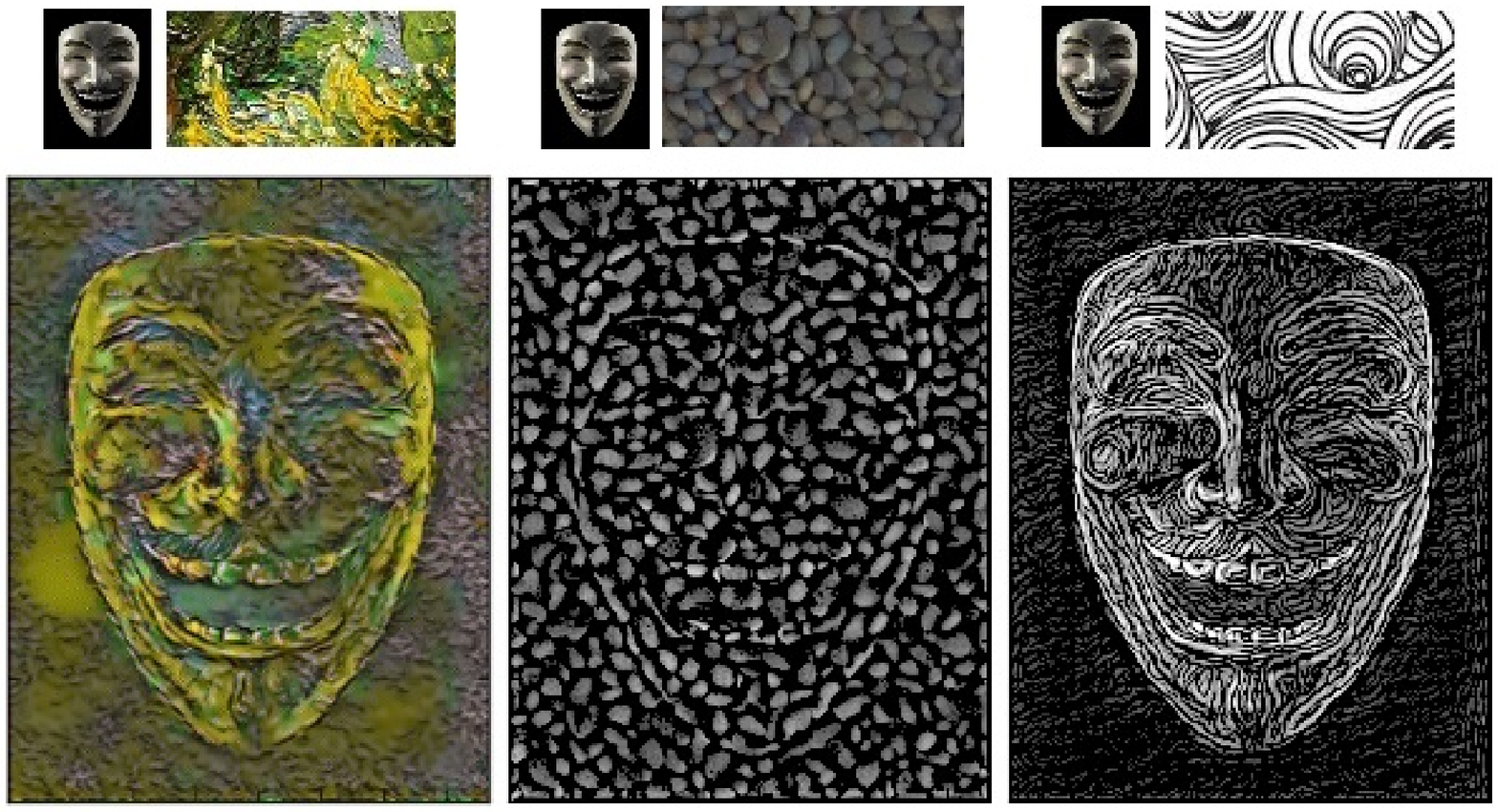}
\epsfxsize=11.5cm
\epsfbox{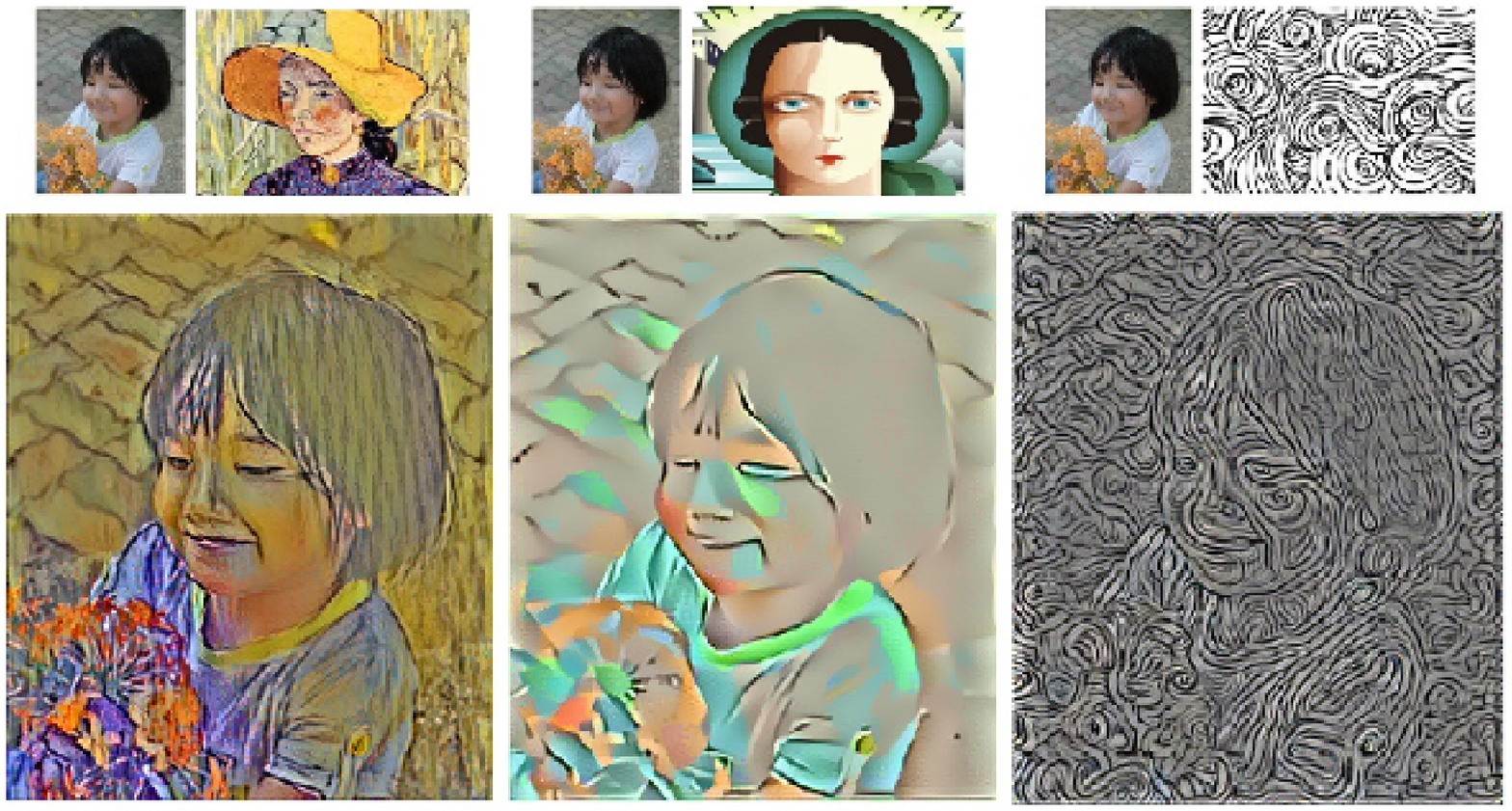}
\end{center}
\caption{Images of a mask (top row) and images of a child (bottom row) are synthesized using combination of the style loss (Eq \ref{style}) and the content loss (Eq \ref{content}).}
\label{resultsynim2}
\end{figure}

An image synthesized using the loss function from equation \ref{content} displays an exact replica of the image since it is a generation based on content loss. Combining the losses from both content loss and style loss, the relationships among components can be obtained via content loss while the texture is obtained via style loss. Figure \ref{resultsynim2} shows synthesized images from equations \ref{style} and \ref{content} which represent content loss and style loss respectively.  Combining weighted losses from both functions produces an interesting output since the pixel information from two different sources is blended together. The blending is not according to the spatial positions of the pixels(as in Eq \ref{blend}) but from a deeper abstraction obtained from hidden layers of a deep neural network. This gives a kind of control known as \emph{style transfer}, where one image provides information on the content and the other image provides information about the style. New synthesized images successfully capture both content and style information. This provides a new interesting generative approach.

\section{Reflection \& Discussion}
In \cite{hertzmann01}, the \emph{image analogies} learn a generative function between a pair of images $g: A \mapsto A'$ . The generative function $g(\cdot)$ learns a specific transformation which can be applied to other images. The transformation is, however, limited to the specific learnt function $g(\cdot)$. Leverage on recent advances in deep learning, pre-trained models (e.g., deep dream, VGG networks) are employed in the generative process. This allows a richer transformation style since the deep neural network acts as a transform function that re-represents information of a given image in the hidden layers.  In \cite{gatys16}, two classes of loss functions: content loss ${\mathcal{L}}_{content}$ and style loss ${\mathcal{L}}_{style}$ are proposed. This allows different combinations of style and content to be realized with ease.
 
We offer a summary of the creative process using distributed representation as follows: let $N^{r \times c}$ and $T^{r \times c}$ be a matrix of input pixels from white noise and the target image $T$ respectively. Feeding  $N^{r \times c}$ and $T^{r \times c}$ to the VGG network  produces two set of activations  ${\bf h}^l_n$ and ${\bf h}^l_t$ in the hidden layers. Gradually reducing the discrepancy between ${\bf h}^l_n$ and ${\bf h}^l_t$ should conceptually synthesize the image based on information from the image $T^{r \times c}$. Let $G_{L}^{r \times c}$ be a gradient matrix computed from the loss function ${\mathcal L}({\bf h}_n,{\bf h}_i)$, hence, the generative model can be expressed as an iterative update:
\begin{equation}
N^{r \times c} \leftarrow N^{r \times c} + \delta G_{L}^{r \times c}
\end{equation}
where $\delta$ is the step size. That is, an image $N$ gradually transforms into a new image using information from $T$. The synthesized image will share many characteristics with the original image depending on the loss functions. 
The content loss is, in essence, the differences between the synthesized image (initialized using white noise) and the target image:
\begin{equation}
{\mathcal L}_c({\bf h}_n,{\bf h}_t) \propto \frac{1}{k_c} \sum_l ({\bf h}_n - {\bf h}_t)_l^2
\end{equation}
where $k_c$ is a constant normalizing the loss. ${\mathcal L}_c$ minimizes one to one relationship between the nodes in the hidden layers and thus preserve original content. On the other hand, the style loss ${\mathcal L}_s$ minimizes the gram matrix in the hidden layers. Minimizing the Gram matrix abstracts away spatial information (since the inner product only correlates the feature map as a whole and not the detail inside the feature map).
\begin{equation}
{\mathcal L}_s({\bf h}_n,{\bf h}_t) \propto \frac{1}{k_s} \sum_l ({\bf h}_n^T{\bf h}_n - {\bf h}_t^T{\bf h}_t)_l^2
\end{equation}
where $k_s$ is a constant normalizing the loss. In \cite{li17}, the authors argue that the essence of style transfer is to match the feature distribution between the style and the generated image and shows that minimizing the gram matrix is equivalent to minimizing the Maximum Mean Discrepancy (MMD) with the second order polynomial kernel.

\section{Conclusion \& Future Work}
The synthesis of an image using the information obtained from distributed VGG layers have many strengths (i) the approach often produces visually appealing images, more appealing than those produced by a filter technique e.g., artistic filters; (ii) the approach offers a flexible means to combine different content images and style images together. The synthesized output convincingly shows that style loss produces an image with a clear local texture but often lacks a clear relationship among texture components over a long spatial distance. Source images having strong local texture such as pebble, line drawing, etc., will produce impressive outcomes.

The issue of long range dependency is a universal issue in all domains and researchers have approached this differently in different domains. For example, a Long Short-Term Memory (LSTM) \cite{hochreiter97} is an enhanced recurrent neural network that has been successfully applied to speech, text and image processing. Combining context loss and style loss to synthesize a new image offers a means to deal with long range dependency issue in images. The approach always produces interesting output since the content loss always preserves the content while the style loss decorates the existing content with the style texture. In future works, we wish to further explore how to assert controls into the generative process \cite{span12,champandard16}.

\begin{small}
\subsubsection*{Acknowledgments}
We would like to thank the GSR office  for their partial financial support given to this research. 
\end{small}

%
%


\begin{thebibliography}{}
\begin{small}
\bibitem{span12}
Phon-Amnuaisuk, S., Panjapornpon, J.:
Controlling generative processes of generative art.
In: Proceedings of the International Neural Network Society Winter Conference (INNS-WC 2012).
Procedia Computer Science 13:43-52 (2012)

\bibitem{nina15}
Mohd Salleh, N.D., Phon-Amnuaisuk, S.:
Quantifying aesthetic beauty through its dimensions: a case study on trochoids.
International Journal of Knowledge Engineering and Soft Data Paradigms 5:51-64 (2015)

\bibitem{mandelbrot83}
Mandelbrot, B.:
The Fractal Geometry of Nature. New York: W.H. Freeman (1983)

\bibitem{lindenmayer90}
Prusinkiewicz, P., Lindenmayer, A.:
The Algorithmic Beauty of Plants. Springer (1990)

\bibitem{schmidhuber15}
Schmidhuber, J.:
Deep learning in neural networks: An overview. 
Neural Networks, 61:85-117. (2015)

\bibitem{lecun15}
LeCun, Y., Bengio, Y., and Hinton, G.: 
Deep learning. Nature, 521:436–444. (2015)

\bibitem{zeiler13}
Zeiler, M.D., Fergus, B.:
Visualizing and Understanding Convolutional Networks.
In: Proceedings of the European Conference on Computer Vision. (ECCV 2014) pp. 818-833 (2013)

\bibitem{zhou17}
Zhou, B., Bau, D., Oliva, A., Torralba, A.: 
Interpreting deep visual representations via network dissection.
arXiv:1711.05611 (2017)


\bibitem{simonyan15}
Simonyan, K., Zisserman, A.: 
Very deep convolutional networks for large-scale image recognition. 
In: Proceedings of the International Conference on Learning Representations (ICLR 2015) arXiv:1409.1556 (2015)

\bibitem{kalajd08}  
Kalajdzievski, S.:
Math and Art: An Introduction to Visual Mathematics.
CRC Press (2008)

\bibitem{strothotte02} 
Strothotte, T., and Schlechtweg, S.:
Non-Photorealistic Computer Graphics: Modeling, Rendering and Animation.
Morgan Kaufmann Publishers, Elsevier Science, USA. (2002)

\bibitem{spanazhan14}
Ahmad, A. Phon-Amnuaisuk, S.:
Emulating pencil sketches from 2D images.
In: Proceedings of the International Conference on Soft Computing and Data Mining  (SCDM 2014) pp. 613-622 (2014)

\bibitem{he16}
He, K., Wang, Y., Hopcroft, J.: 
A powerful generative model using random weights for the deep image representation. 
In: Proceedings of the Annual Conference on Neural Information Processing Systems (NIPS 2016), Barcelona, Spain. pp. xxx-xxx (2016)

\bibitem{hertzmann01}
Hertzmann, A., Jacobs, C.E., Oliver. N., Curless, B., Salesin, D.H.
Image Analogies.
In: Proceedings of the 28th Annual Conference on Computer Graphics and Interactive Techniques.  ACM Press / ACM SIGGRAPH. pp 327–340 (2001)

\bibitem{gatys16}
Gatys, L.A., Ecker, A.S., Bethge, M.: 
Image style transfer using convolutional neural networks. 
In: Proceedings of the IEEE Conference on Computer Vision and Pattern Recognition (CVPR 2016), pp. 2414-2423 (2016)

%
\bibitem{wolfram84}
Wolfram, S.: 
Cellular automata as models of complexity. 
Nature 331(4) 419-424 (1984)

%
\bibitem{mahendran15}
Mahendran, A., Vedaldi, A.:
Understanding deep image representations by inverting them.
In: Proceedings of the IEEE Conference on Computer Vision and Pattern Recognition (CVPR 2015) pp. 5188-5196 (2015)

\bibitem{gatys15}
Gatys, L.A., Ecker, A.S., Bethge, M.: 
Texture synthesis using convolutional neural networks. 
In: Proceedings of of the Annual Conference on Neural Information Processing Systems (NIPS 2015) 262-270 (2015)

\bibitem{li17}
Li, Y., Wang, N., Liu, J., Hou, X.: 
Demystifying neural style transfer
arXiv:1701.01036v2 (2017)

\bibitem{hochreiter97}
Hochreiter, S., Schmidhuber, J.:
Long short-term memory.
Neural Computation 9(8):1735-1780 (1997)

\bibitem{champandard16}
Champandard, A.J.: 
Semantic style transfer and turning two-bit doodles into fine artwork
nuci.ai Conference 2016, Artificial Intelligence in Creative Industries. arXiv:1603.01768v1 (2016)


\end{small}
\end{thebibliography}
\end{document}